\documentclass[11pt, a4paper]{article}
\usepackage[utf8]{inputenc}
\usepackage[a4paper, margin=1in]{geometry}
\usepackage{graphics}
\usepackage{authblk}
\usepackage{graphicx}
\usepackage{amsfonts}
\usepackage{amsmath}
\usepackage{amssymb}
\usepackage{multirow}
\usepackage{array}

\usepackage{caption}
\usepackage[subtle]{savetrees}
\usepackage{titlesec}
\usepackage{textcomp}
\usepackage{enumerate}
\usepackage[shortlabels]{enumitem}

\begin{document}


\title{ITRUSST Consensus on Standardised Reporting for Transcranial Ultrasound Stimulation}
\author[1,2]{Eleanor Martin}
\author[3]{Jean-François Aubry}
\author[4]{Mark Schafer}
\author[5]{Lennart Verhagen}
\author[1,*]{Bradley Treeby}
\author[6,*]{Kim Butts Pauly}

\affil[1]{\raggedright
Department of Medical Physics and Biomedical Engineering,\newline University College London, London, U.K}
\affil[2]{Wellcome/EPSRC Centre for Interventional and Surgical Sciences,\newline University College London, London, UK}
\affil[3]{Physics for Medicine Paris, Inserm U1273, ESPCI Paris, CNRS UMR8063, \newline PSL University, Paris, France}
\affil[4]{School of Biomedical Engineering, Science and Health Systems, \newline Drexel University, Philadelphia, PA USA}
\affil[5]{Donders Institute for Brain, Cognition and Behaviour, \newline Radboud University, 6525 GD Nijmegen, the Netherlands}
\affil[6]{Department of Radiology, \newline Stanford University, Stanford, CA, USA
\endgraf \textup{ * These authors contributed equally.
        }
        
        \textup{ Corresponding Author: Kim Butts Pauly: kimbutts@stanford.edu
        }}

\date{Feb 2024}

\maketitle

\interfootnotelinepenalty=10000


\begin{abstract}
As transcranial ultrasound stimulation (TUS) advances as a precise, non-invasive neuromodulatory method, there is a need for consistent reporting standards to enable comparison and reproducibility across studies. To this end, the International Transcranial Ultrasonic Stimulation Safety and Standards Consortium (ITRUSST) formed a subcommittee 
of experts across several domains
to review and suggest standardised reporting parameters for low intensity TUS, resulting in the guide presented here. 
The scope of the guide is limited to reporting the ultrasound aspects of a study. The guide
and supplementary material
provide a simple checklist covering the reporting of: (1) the transducer and drive system, (2) the drive system settings, (3) the free field acoustic parameters, (4) the pulse timing parameters, (5) \emph{in situ} estimates of exposure parameters in the brain, and (6) intensity parameters. Detailed explanations for each of the parameters, including discussions on assumptions, measurements, and calculations, are also provided.
\end{abstract}

\


\clearpage

\section{Introduction}

Transcranial ultrasound stimulation (TUS) is a non-invasive neuromodulation technique that employs focused ultrasound waves to modulate neuronal activity within the brain. TUS offers a promising avenue for therapeutic and research applications due to its spatial precision and ability to target deep neural structures. As the field transitions to more widespread human studies, the time is upon us to standardize the reporting of such studies to aid understanding and reproducibility.

To this end, the International Transcranial Ultrasonic Stimulation Safety and Standards Consortium (ITRUSST) formed a subcommittee to review and suggest standardised reporting parameters for low intensity TUS, with multiple presentations to the ITRUSST group and multiple opportunities for feedback, resulting in the guide presented here. Similar guidelines have been produced for other analogous techniques and applications \cite{ter2011,padilla2022}.

This guide focuses on the ultrasound aspects of TUS experiments. It does not discuss reporting of additional elements such as EEG, fMRI, or behavioral readouts, which are important, but outside the scope of this work. It does not seek to establish safety values or thresholds, which are addressed in a separate ITRUSST consensus on TUS safety \cite{aubry2023itrusst}. While this guide endeavors to align with existing international standards related to diagnostic ultrasound imaging and focused ultrasound transducers, it may diverge from these standards when necessary. An important additional note is that we make the assumption that the ultrasound fields that this guide applies to are linear.  Our rationale for doing so is provided in Sec. \ref{Sec:App_linearity} of the Supplementary Material. 

This guide is structured as follows. In Section \ref{Sec:freefieldparams}, we discuss reporting of the transducer and drive system, and the free field acoustic pressure amplitude and spatial characteristics of the focal region, as measured by a hydrophone in free field in a water bath. These measurements, performed under standardised conditions, serve as a baseline for comparison between studies independent from the specific configuration of use during TUS studies.  In Section \ref{Sec:timingparams}, we discuss the pulse timing parameters. Next, in Section \ref{Sec:derivedparams}, we discuss the reporting of exposure parameters, which describe the acoustic field inside the individual brain, after accounting for the skull bone, brain tissue, and any other acoustic distortions. Further details and a summary checklist of all required reporting parameters are provided in the Supplementary Material, Sections \ref{Sec:App_linearity}-\ref{Sec:App_checklist}.


\section{System and Free Field Acoustic Pressure Parameters}
\label{Sec:freefieldparams}

System parameters describe the type and geometry of the transducer, and the signal chain or system used to drive the transducer in order to generate the acoustic field. The free field acoustic pressure parameters provide a description of the generated ultrasound field under reference conditions. The term ‘free field’ refers to the field generated when a transducer is radiating continuously \cite{international1iec} (or with the comparatively long pulses used in TUS studies) into water without any obstruction by reflectors, scatterers, or aberrators. See Sec. \ref{AppA} for details of how this is implemented during measurements.

Reporting  free field acoustic parameters allows comparison of the transducer output at the chosen output level and focal settings under standardised conditions, serving as a baseline for comparison between studies. In this guide, the term output level refers to the amplitude of the acoustic pressure output, which is governed by settings on the drive system such as voltage, power or intensity. The focal position setting refers to the position of the focal region where this can be steered to different positions. It may be fixed for a particular transducer, or set by the operator, for example, by selecting a distance setting, or by setting the relative phases applied to the transducer elements.  The free field parameters described in this section should be reported at the output and focal settings used during exposure of participants during studies.

\subsection{Transducer and Drive System}

\subsubsection{Transducer Description} \label{Sec:tx_desc}
A description of the ultrasound transducer should be given, including the manufacturer and model number, and the operating frequency of the transducer. For spherically focussing transducers, the geometry is described by the radius of curvature and aperture diameter (see Fig. \ref{fig:transducers}). In some cases, the exact geometry may not be known, for example, if the transducer construction includes a lens or permanently attached coupling medium, in which case, any other information such as the nominal position of the focus relative to the transducer face should be reported. For multi-element transducers, the number of elements and information about their shape, size, and positions should be given if available. The exact size and position of elements may not be available for commercial transducers, so a brief description accompanied by the model number is sufficient. Any other features of the transducer such as permanently attached coupling media, or lenses should be described, including a description of the material, its thickness and shape/geometry, and properties of the material such as the sound speed, density and attenuation coefficient where known.

\subsubsection{Drive System Description} \label{Sec:drive_system_desc}
The input signal used to drive the transducer may be generated by an integrated drive system, or using a signal generator and RF amplifier. All components of the system, including manufacturer and model numbers and any external electrical impedance matching networks used to couple the drive signal to the transducer should be reported.  

\subsubsection{Drive System Settings} \label{Sec:drive_system_settings}
All drive system settings used during studies should be described, including the operating frequency, output level settings (e.g. displayed electrical power, focal intensity, voltage etc.), and for multielement transducers, focal distance or position settings applied. 
The method of coupling the transducer to the participant forms part of the \emph{in situ} acoustic transmission path, so the materials and methods used should be described.

\begin{figure}
    \centering
    \includegraphics[width=\textwidth]{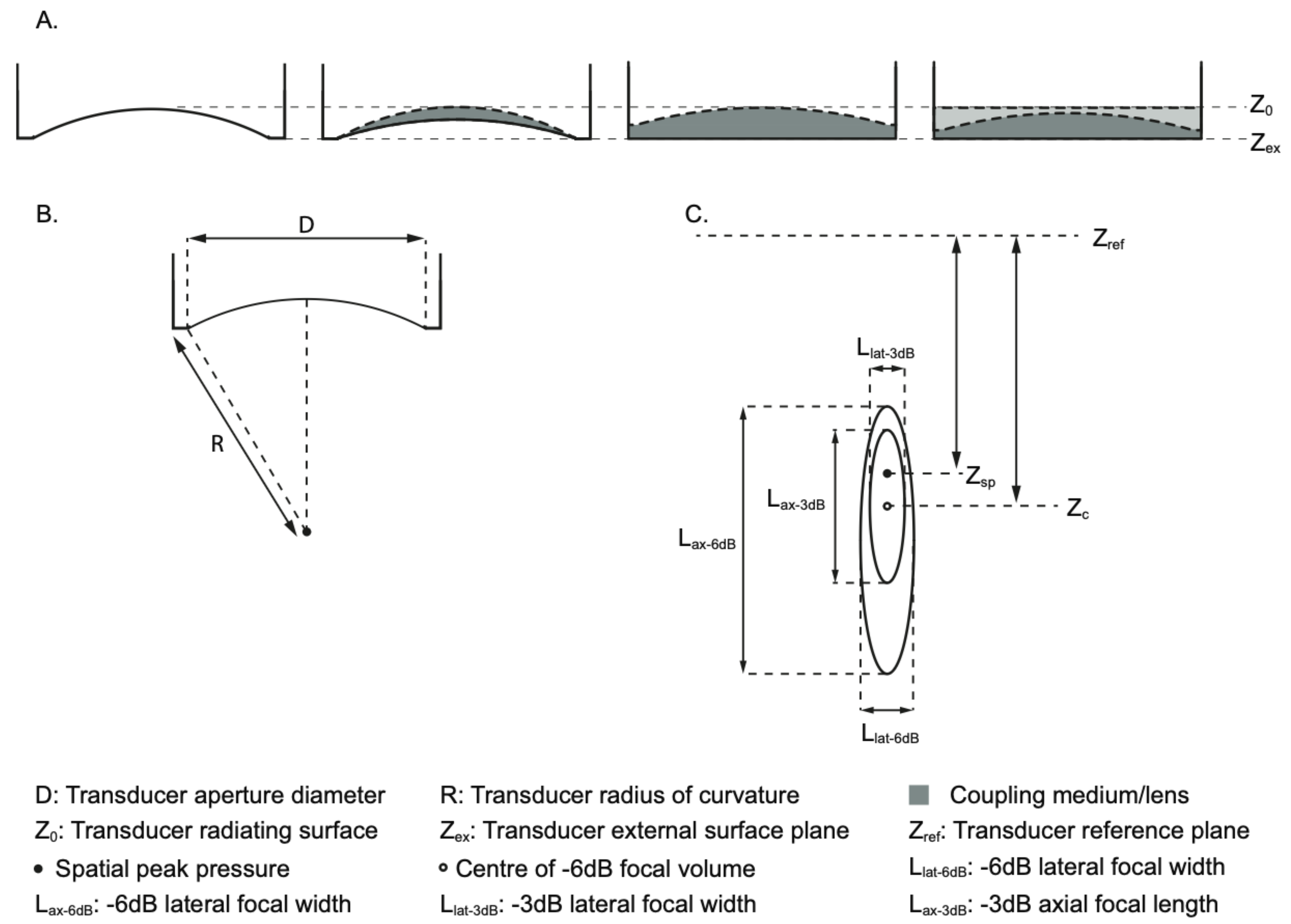}
    \caption{Schematics of example transducers, geometric and field parameters, for information. (A) common transducer configurations including attached coupling media or internal lenses. The external transducer surface plane, $Z_{ex}$ which may be coincident with the front surface of the transducer housing often serves as the reference plane, $Z_{ref}$ from which distances are measured. For some transducers, e.g. complex multi-element devices, $Z_{ref}$ may intersect with the centre of the surface on which the elements sit. (B) Aperture diameter, D, and radius of curvature, R, are used to describe the geometry of focusing transducers. (C) Free field pressure parameters shown with respect to the reference plane. $Z_{sp}$ is the axial position of the location of spatial-peak pressure relative to the reference plane $Z_{ref}$; $Z_c$ is the axial position of the centre of the -3dB focal region relative to the reference plane.}
    \label{fig:transducers}
\end{figure}

\subsection{Free Field Acoustic Parameters}

\subsubsection{Parameters To Report} \label{Sec:ff_param}
The free field acoustic pressure field parameters are illustrated in Fig. \ref{fig:transducers}. Table 1.1 shows the quantities to be reported which includes spatial peak pressure amplitude, focal dimensions, and location.  
An  explanation of how to calculate focal dimensions from the axial pressure profile is given in Sec. \ref{dB}. Two illustrative reporting examples are provided in Table 1.1, and annotated example axial pressure profiles are shown in Fig. \ref{fig:A2}. 

The amplitude of the spatial-peak acoustic pressure should be reported at each of the output level settings, and all quantities should be reported for each of the focal position settings used during studies, where practical. This will provide a reference spatial-peak pressure in water for each study exposure. For multi-element array devices where steering and aberration correction are performed, the focal position and amplitude may vary between participants and target locations. Therefore, reporting of free field parameters for each condition may not be useful. Instead information about the pressure amplitude and size of the focal region over the focal steering range utilised during the study should be reported. See Sec. \ref{Sec:av_params} for recommendations for reporting average or ranges of parameters.

\begin{table}
\textbf{A}\\
\textbf{Transducer and Drive System Parameters} \\
\footnotesize
\begin{tabular}{>{\raggedright}m{1.7cm} m{2.8cm} m{1.4cm} m{1.4cm} m{1.3cm} m{1.3cm} m{2.8cm}  }
 & Manufacturer, Model Number  & Centre Frequency  & Radius of Curvature  & Aperture Diameter  & Number Elements & Element Distribution \\ \cline{2-7} 
Transducer  &  \multicolumn{1}{|p{2.8cm}}{TX1, Manufacturer 1}   & 250 kHz  & 64 mm   & 64 mm    & 2   & Spherical cap, annular array, equal area \\
Matching  & \multicolumn{4}{|l}{Electrical impedance matching network, Manufacturer 1}   &  &   \\
Integrated drive system & \multicolumn{3}{|l}{2-channel drive system, Manufacturer 1} &  &  &       \\

\end{tabular}
\newline 
\newline 

\normalsize
 \textbf{Drive System Settings} \\
\footnotesize
\begin{tabular}{m{1.7cm} m{2.8cm} m{2.8cm} m{3cm}}
  & Operating Frequency & Output level Setting & Focal Position Setting \\ \cline{2-4} 
Experiment 1-4 & \multicolumn{1}{|p{2.8cm}}{270 kHz}   & 4.78 W     & Phases: 0, 66.3º \\

\end{tabular}
\newline 
\newline 

\normalsize
\textbf{Free Field Pressure Parameters} \\
\scriptsize
\begin{tabular}{m{1.7cm} m{1.8cm} m{1.8cm} m{1.9cm} m{1.2cm} m{1.2cm} m{1.2cm} m{1.2cm} }
  & Spatial Peak Pressure Amplitude & Axial Position Spatial Peak   Pressure & Position of Centre of Axial -3dB Pressure & Axial -3dB Size & Lateral -3dB Size & Axial -6dB Size & Lateral -6dB Size \\ \cline{2-8} 
\footnotesize Experiment 1-4 & \multicolumn{1}{|p{1.6cm}}{\footnotesize 700 kPa}   & \footnotesize 43 mm   & \footnotesize 48 mm   & \footnotesize 30 mm    & \footnotesize 5 mm    &\footnotesize  48.7 mm   & \footnotesize 7 mm   \\ 
 & \multicolumn{7}{l}{\footnotesize Notes: Reference plane is coincident with the centre of the radiating surface} \\

\end{tabular}
 \newline 
  \newline 

\normalsize
\textbf{B}\\
\textbf{Transducer and Drive System Parameters} \\
\footnotesize
\begin{tabular}{>{\raggedright}m{1.7cm} m{2.8cm} m{1.4cm} m{1.4cm} m{1.3cm} m{1.3cm} m{2.8cm}  }
 & Manufacturer, Model Number  & Centre Frequency  & Radius of Curvature  & Aperture Diameter  & Number Elements & Element Distribution \\ \cline{2-7} 
Transducer  &  \multicolumn{1}{|p{2.8cm}}{TX2, Manufacturer 2}   & 650 kHz  & 80 mm   & 61 mm    & 1   & Spherical cap \\
Integrated drive system  & \multicolumn{4}{|l}{Ultrasound driving systems, Manufacturer 2}   &  &   \\   

\end{tabular}
\newline  
\newline  

\normalsize
 \textbf{Drive System Settings} \\
\footnotesize
\begin{tabular}{m{1.7cm} m{2.8cm} m{2.8cm} m{3cm}}
  & Operating Frequency & Output level Setting & Focal Position Setting \\ \cline{2-4} 
Experiment 1 & \multicolumn{1}{|p{2.8cm}}{650 kHz}   & $p_r$ = 0.72 MPa
     & fixed  \\

\end{tabular}
     \newline 
          \newline

\normalsize
\textbf{Free Field Pressure Parameters} \\
\scriptsize
\begin{tabular}{m{1.7cm} m{1.8cm} m{1.8cm} m{1.9cm} m{1.2cm} m{1.2cm} m{1.2cm} m{1.2cm} }
 & Spatial Peak Pressure Amplitude & Axial Position Spatial Peak   Pressure & Position of Centre of Axial -3dB Pressure & Axial -3dB Size & Lateral -3dB Size & Axial -6dB Size & Lateral -6dB Size \\ \cline{2-8} 
\footnotesize Experiment 1 & \multicolumn{1}{|p{1.6cm}}{\footnotesize 720 kPa}  &   & \footnotesize 80 mm   &    &   & \footnotesize 30 mm  & \footnotesize 5 mm    \\    
 & \multicolumn{7}{l}{\footnotesize Notes:  Reference plane is coincident with the outer surface of the transducer housing} \\

\end{tabular}
 \newline 
  \newline 
\captionsetup{labelformat=empty}
 \caption{Table 1.1: Transducer and drive system description, drive system settings, and free-field pressure parameters adapted from Johnstone et al \cite{johnstone} and Badran et al. \cite{Badran2020}. }
   \label{fig:table1}
\end{table}

\subsubsection{Methods of Obtaining Free Field Pressure Parameters}

The free field pressure parameters can be obtained by investigators in one of several ways, for example, directly from acoustic field measurements using a hydrophone in a water bath, from a test report provided by the manufacturer or other calibration/characterisation provider, or via a combination of simulation and measurement. In any case, the source and/or method used to obtain the field parameters should be reported, and where measurement or simulation were used, the procedures should be described, as well as the equipment used (including hydrophone type and element size, and calibration parameters). Further guidance can be found in Sec. \ref{AppA}. 

For those performing their own measurements to determine the free field acoustic pressure parameters, it is recommended that information on measurement best practice be sought from hydrophone measurement standards \cite{IEC62127, IEC61828, international1iec} with additional help from the literature (e.g. \cite{Harris2022, OutputMeas, Martin2019}). While a full description is out of scope of this document, some further details and discussion are provided in Sec. \ref{AppA}. Similarly, where simulation is used to obtain some or all of the field parameters, the literature should be consulted for methods and best practice. 

\subsubsection{Spatial-Peak Pressure at Study Output Levels}
The free field spatial-peak pressure amplitude at each of the study output level settings and focal position settings may be obtained either directly from a measurement with the transducer operated at the study output level and focal setting, or by scaling a measurement made with the same focal setting, but at a different output level. Typically, a lower output level can be used for free field measurements than would be used to obtain a similar pressure amplitude in the brain after propagation through the skull. Measurements with sufficient signal to noise ratio can be made, while reducing the possibility of causing damage to the hydrophone (especially at the lowest frequencies). This is discussed in more detail in Sec. \ref{AppA}. 

\subsubsection{Defining the Spatial Location of Reported Free Field Pressure Parameters} 
The spatial location of the spatial-peak pressure amplitude can be defined relative to a variety of different reference locations, for example the radiating surface or the external transducer surface plane, also known as the exit plane (see Fig. \ref{fig:transducers}). In some cases the full construction of the transducer or the distance from the radiating surface to the outer surface of any permanently attached or integrated lens or coupling medium may not be known. In this case, a useful reference location would be the outer surface of these layers, the external transducer surface plane. 
The chosen reference plane location should be reported along with the parameters. 

\subsubsection{Uncertainty on Free-Field Acoustic Parameters}

All hydrophone measurements have an associated measurement uncertainty which arises from a combination of systematic and random uncertainties. One major source of systematic uncertainty is hydrophone sensitivity, which can be up to 20\% depending on frequency, and calibration method. Additionally, uncertainty on pressure measurements will propagate to derived quantities. Understanding of uncertainty on pressure measurements is essential when reporting and comparing measurements. Further discussion of this can be found in Sec \ref{Sec:UC}.


\section{Pulse Timing Parameters}
\label{Sec:timingparams}

Typical pulse timing parameters are laid out in Figure 2.The period refers to the duration of one cycle of the operating frequency $f_0$. A pulse is a single continuous sonication and has a duration referred to as the pulse duration (PD). If the pulse is repeated, the pulse repetition interval (PRI) is the time between successive pulses. The pulse repetition frequency is given by $PRF=\frac{1}{PRI}$. 

\begin{figure}
    \centering
    \includegraphics[width=\textwidth]{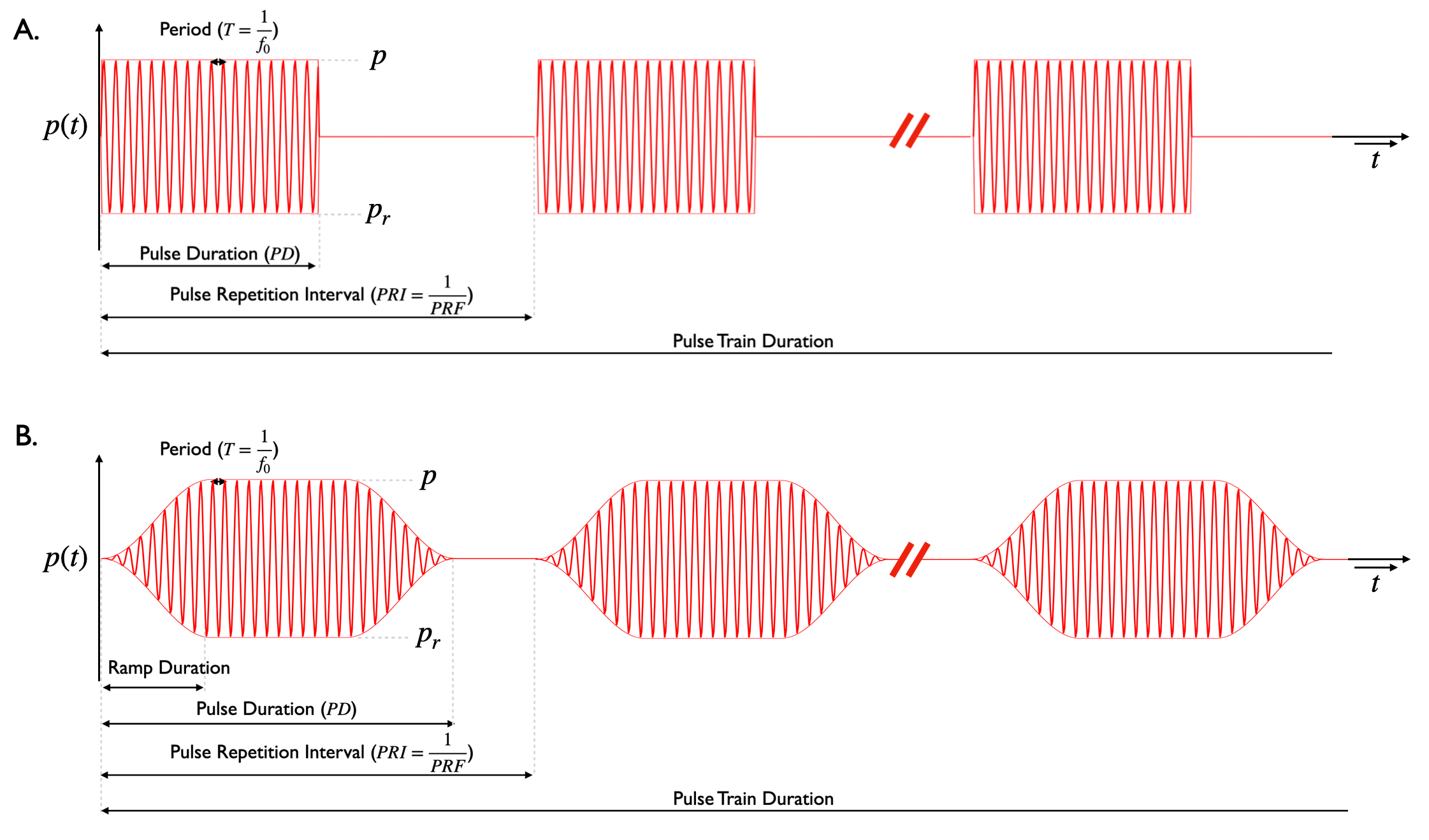}
  \captionsetup{labelformat=empty}    
  \caption{Figure 2. Schematics of two  ultrasound pressure waveforms. A  rectangular ramp shape is used in A, while a Tukey window on the pressure waveform is used in B. The pressure amplitude is shown as $p$, while the peak negative pressure is shown as $p_r$; these should be the same when operating at low pressures within the linear regime. The duration of a single cycle of the operating frequency is the period (T). A single continuous sonication is a pulse with a duration of pulse duration (PD). Pulses are often repeated in a pulse train. The time between two pulses in a pulse train is the pulse repetition interval (PRI) and is equal to 1 divided by the pulse repetition frequency (PRF). The pulse train has a duration which is the pulse train duration. The pulse train can be repeated, and if so has a structure similar to the pulse, as laid out in Tables 2.1-2.3. In this figure, the operating frequency $f_0$ is lower than typically used in order to aid visualization.}
    \label{fig:timing}
\end{figure}

In Tables 2.1-2.3 we suggest a table reporting structure. In the rare case of only a single continuous pulse, only the first line of a table need be provided, with the pulse duration, and if a ramp is used, the pulse ramp duration and the pulse ramp shape. The pulse repetition interval is not applicable. When the pulse is repeated, the PRI/PRF are entered in the first line, and the second line provides information on the pulse train including the pulse train duration, the pulse train ramp duration, and the pulse train ramp shape.

The examples chosen from the literature illustrate a number of variations in TUS pulse parameters in order to illustrate increasing levels of patterning, as well as rectangular and ramped waveforms. While figures are helpful and illustrative, we feel that tables of the pulse parameters would be most clear. 

\begin{table}
\begin{tabular}{  m{.1cm}  m{1.2cm} |m{1.8cm} m{1.5cm} m{1.5cm} m{2.4cm} m{4cm} } 
& & \_Duration & \_Ramp Duration & \_Ramp Shape & \_Repetition Interval/ Frequency & Notes \\   
  \hline
\textbf{A}& Pulse\_ & {\small 2 s} & {\small 0} & {\small rectangular} & {\small 4 ms/250 Hz} & \\ [1ex] 
& Pulse Train\_ & {\small 0.3 s} &{\small 0} & {\small rectangular} &  & \\ [1ex] 
  \hline
\textbf{B}& Pulse\_ & {\small 3.25 ms} & 1 ms & {\small Tukey on Pressure} & {\small 4 ms/250 Hz} & \\ [1ex] 
& Pulse Train\_ & {\small 0.3 s} & {\small 0} & {\small rectangular} & & {\small A 1s 250Hz square wave auditory mask was delivered using cicumaural headphones. The mask was synchronised to start 100 ms before the onset of TUS pulse trains.} \\ [1ex] 
   
\end{tabular}
    \captionsetup{labelformat=empty}
   \caption{Table 2.1: Pulse timing parameters from two experiments in Johnstone \emph{et al.} \cite{johnstone}. The A stimulation is experiment 1, condition A2 and the B stimulation is experiment 4, condition A2. In both cases, the first line contains the pulse characteristics and the second line contains the pulse train characteristics. }
     \label{fig:table2.1}
\end{table}

A number of phrases are commonly used to refer to ultrasound pulse trains. We recommend avoiding ambiguous phrases while retaining clear conventions of the community. The use of the word “burst” is ambiguous as it has different meanings in different contexts. For example, in diagnostic ultrasound, burst refers to a single long pulse, while on function generators and in TMS, burst is often used to refer to pulse trains. We suggest avoiding the use of the word “burst” to remove this ambiguity and instead to use the terms “pulse” and “pulse trains."

The phrase ``repetitive TUS" (rTUS) is conventionally used to refer to pulse trains intended to elicit cumulative or delayed effects, as opposed to only acute effects. The distinction between acute and delayed effects is critical for study design (`online' vs. `offline' designs), the underlying neurophysiology (acute modulatory vs. early-phase plasticity mechanisms), and the safety assessment. Distinguishing characteristics of an rTUS protocol, for example, the pulse repetition frequency, can be included in the label, as in the phrase `10 Hz rTUS'. It is recommended to avoid using unique labels when an (extended) rTUS label is also appropriate. By definition, all rTUS protocols include at least the pulse and pulse train timing parameters, as in the 5 Hz rTUS example A of Table 2.2.  Patterned rTUS protocols are a subset of rTUS where the pulse trains are repeated in a pattern and, thus, include more than two rows, as in the patterned rTUS example B of Table 2.2.

\begin{table}
\begin{tabular}{  m{.5cm}  m{2cm} |m{2cm} m{2cm} m{2cm} m{2.3cm} m{1cm} } 
& & \_Duration & \_Ramp Duration & \_Ramp Shape & \_Repetition Interval/ Frequency & Notes \\   
  \hline
\textbf{A}& Pulse\_ & {\small 20 ms} & {\small 0} & {\small rectangular} & {\small 200 ms/5 Hz} & \\ [1ex] 
& Pulse Train\_ & {\small 80 ms} & {\small 0} & {\small rectangular} &  & \\ [1ex] 
  \hline
\textbf{B}& Pulse\_ & {\small 0.32 ms} & {\small 0} & {\small rectangular} & {\small 1 ms/1 kHz} & \\ [1ex] 
& Pulse Train\_ & {\small 500 ms} & {\small 0} & {\small rectangular} & {\small 1.6 s/0.625 Hz} & \\ [1ex] 
&  Pulse Train Repeat\_ & {\small 80 s} & {\small 0} & {\small rectangular} & & \\ [1ex] 
   
\end{tabular}
    \captionsetup{labelformat=empty}
    \caption{Table 2.2. Pulse timing parameters for the two stimulations from Zeng \emph{et al.} \cite{zeng}. A was a single pulse train repeated every 200 ms for 80 s. B consisted of a short pulse train repeated every 1.6 s, again for a total time of 80 s. This is an example of patterned repetitive TUS. }
    \label{fig:table2.2}
\end{table}

\begin{table}
\begin{tabular}{ m{1.6cm} |m{1.8cm} m{1.8cm} m{2cm} m{2.3cm} m{4cm} } 
 & \_Duration & \_Ramp Duration & \_Ramp Shape & \_Repetition Interval/ Frequency & Notes \\   
  \hline
 Pulse\_ & {\small 0.5 ms} & {\small 0} & {\small rectangular} & {\small 1 ms/1 kHz} & \\ [1ex] 
 Pulse Train\_ & {\small 0.3 s} & {\small 0} & {\small rectangular} & {\small 1 s/1 Hz} & \\ [1ex] 
  Pulse Train Repeat\_ & {\small 30 s} & {\small 0} & {\small rectangular} & {\small 60 s/0.017 Hz} & {\small TUS and Light+TUS conditions were interleaved with Light Only and No Stimulus conditions}\\ [1ex] 
   Repeat 2\_ & {\small 40 min} & {\small 0} & {\small rectangular} &  & \\ [1ex] 
\end{tabular}
   \captionsetup{labelformat=empty}
    \caption{Table 2.3. Pulse timing parameters from Gaur \emph{et al.}\ \cite{gaur} and Mohammadjavadi \emph{et al.} \cite{mohammadjavadi2022}. Pulse trains had two levels of repetitions. }
    \label{fig:table2.3}
\end{table}

For rectangular pulses in a single pulse train, duty cycle (DC) is defined as the percentage of time that a pulse is on,

\begin{equation}
    DC=\frac{PD}{PRI}100\%.
    \label{eq:dc_pulse_train}
\end{equation}

\noindent Duty cycle can be optionally reported since it can be easily derived from the timing tables and is only defined for rectangular pulses. However, if provided, we recommend referring to it in a particular way. The rTUS stimulation protocol of Table 2.2A has a single duty cycle of 10\%. In this case, the overall DC is equal to the $DC_{\text{pulse train}}$. However, when rTUS is patterned, such as in the stimulation of Table 2.2B, it would be clearer to refer to the duty cycle of each level of patterning, such as $DC_\text{pulse train}$ and $DC_\text{pulse train repeat}$. The overall $DC$ is then the product of each $DC$. For the patterned rTUS stimulation in Table 2.2B, $DC_\text{pulse train}=32\%,$  $DC_\text{pulse train repeat}=31.25\%$, and the product is the overall duty cycle, $DC=10\%.$ 

Note, our definition of the pulse duration differs slightly from the diagnostic ultrasound definition, which defines it as “1.25 times the interval between the time when the time integral of the square of the instantaneous acoustic pressure reaches 10\% and 90\% of its final value” \cite{IEC62127}. This difference is negligible for long pulses.

While the community is only starting to embrace ramping as a means to reduce the auditory confound \cite{johnstone,mohammadjavadi2019,deffieux2013low}, it is likely that this will continue and potentially become a source of confusion unless reported in a standardised way. The shape and duration of the ramp must be reported as already described, as well as whether that shape is applied to the pressure (voltage) waveform or the intensity (power) waveform. In Table 2.1, we specify how ramping is applied in Johnstone \emph{et al.}\ \cite{johnstone}.

It is likely that the field will continue to evolve in ways that we cannot now anticipate. The table structure allows for evolution by adding columns or rows as needed. 


\section{Derived Parameters: \emph{In Situ} Estimates of Exposure Parameters}

Exposure parameters describe the properties of the acoustic field inside the individual brain, after accounting for the skull bone, brain tissue, and any other acoustic distortions. The exact ultrasound exposure is generally difficult to directly measure \emph{in vivo}. Instead, exposure parameters are typically estimated using knowledge of the free field parameters along with a derating procedure, simulations, or in-direct measurements. Here, \emph{in situ} estimate is used as a general term to describe the procedure used to calculate the exposure parameters, i.e., the \emph{in vivo} or \emph{in situ} properties of the acoustic field. The recommended \emph{in situ} parameters are provided here, with examples provided in Sec \ref{Sec:ex_insitu}.

\label{Sec:derivedparams}
\subsection{Estimated \emph{In Situ} Pressure Amplitude}
\label{Sec:exposureparams}

The skull bone has very different acoustic material properties (sound speed, density, and attenuation) to the surrounding soft tissues. As ultrasound traverses the skull, these differences cause the waves to become distorted and lose energy. Brain tissue also has a higher attenuation coefficient compared to water. Consequently, the acoustic focus inside the brain will generally have a much lower intensity (sometimes by a factor of 10 \cite{albelda2019experimental}). In some cases, the shape and position of the focus can also become distorted. To complicate matters, the shape and properties of the skull vary significantly both within and between subjects. The same system and pulse timing parameters can therefore generate very different acoustic fields in the brain depending on the subject and the position of the transducer. An estimate of \emph{in situ} pressure amplitude should thus be reported. Several different approaches for obtaining this are outlined below. In all cases, the method used should be reported, along with details of any calculations, parameters, and assumptions.

\begin{enumerate}[(a)]

\item \textbf{Simulating the intracranial ultrasound field.} Simulations are increasingly used to calculate the ultrasound and temperature fields inside the skull and brain. The material properties for the simulations might be based on individualised images for each subject or a CT or MR template image.

If using simulations to calculate the exposure parameters, the following should be reported:
\begin{itemize}
    \item Details of the simulation tool.
    \item Details of how the acoustic and thermal material properties are assigned.
    \item How the transducer modelling and positioning was performed within the simulation.    
    \item Details of relevant simulation input parameters, program settings including mesh or grid parameters, and processing steps.
\end{itemize}
An excellent reference to follow is the \emph{Reporting of Computational Modeling Studies in Medical Device Submissions} FDA guidance document \cite{FDAModelling}. Examples from the literature can be found in e.g., \cite{verhagen2019offline} and \cite{lee2016transcranial}.

\item \textbf{Derating.} The simplest approach to calculating exposure parameters is to derate the pressure by applying acoustic attenuation coefficients to the free field parameters. Skull and frequency-specific acoustic attenuation coefficients can be included to give an estimate of attenuation by the skull under different conditions \cite{ATTALI202348}.

If using attenuation factors to calculate the \emph{in situ} exposure parameters, the following should be reported:
\begin{itemize}
    \item  The individual acoustic attenuation coefficients in dB.(MHz.cm)$^{-1}$ or dB.cm$^{-1}$ (for example, for skull, scalp and brain), and the values used for frequency and distance.
    \item  The total attenuation applied in dB.
\end{itemize}
If experimental measurements are used to determine derating factors, for example using a sample of human skulls, the methods, assumptions, and variability should be reported \cite{aubry2023itrusst}.

\item \textbf{Measuring the effect of the intracranial ultrasound field.} In some cases, indirect measurement of the physical effects of the ultrasound field on the brain can be made. For example, a temperature rise induced by the absorption of ultrasound energy could be measured using MR thermometry. Similarly, for some beam shapes, the bulk displacement induced by an acoustic radiation force could be measured using MR-ARFI. With some restrictions, these measurements could be correlated with the ultrasound exposure parameters in the brain \cite{li2022improving}.

If using measurements to calculate the exposure parameters, the following should be reported:
\begin{itemize}
    \item Details of the measured quantity and the measurement method.
    \item Details of how the exposure parameters are calculated from the measured quantity.
\end{itemize}

\end{enumerate}

In principle, all of the spatial acoustic parameters described in Section \ref{Sec:freefieldparams} could be reported after accounting for the skull and brain. In particular, if using simulations to calculate exposure parameters, it is possible to extract the focal characteristics as well as pressure values at the spatial-peak and target locations \cite{aubry2022benchmark}. As a minimum we recommend the following parameters are reported: (1) the \emph{in situ} estimate for the spatial-peak pressure amplitude and its location, and (2) the \emph{in situ} estimate for the pressure amplitude at the target, if it is known and different from the spatial-peak pressure amplitude. The former values are related to safety, the latter values are related to the study efficacy.

\subsection{Estimate of \emph{In Situ} Mechanical Index} \label{Sec:MI}
For diagnostic ultrasound imaging, the mechanical index ($MI$) provides a standardised indicator related to the potential for mechanical bioeffects, specifically cavitation. In lieu of a safety standard specifically related to TUS, $MI$ is also often reported in TUS studies in relation to regulatory limits given in the FDA guidance document \cite{FDAguidance}. A detailed discussion can be found in \cite{IEC62359}. The \emph{in situ} estimate of mechanical index is easily derived from the \emph{in situ} estimate of spatial-peak pressure. 

The mechanical index is formally defined as 
\begin{equation}
    MI = \frac{p_{r, .3}}{ \sqrt{f_0} },
\end{equation}
\noindent where $p_{r,.3}$ is the peak-rarefactional pressure in MPa derated using an attenuation coefficient of 0.3 dB cm$^{-1}$ MHz$^{-1}$, and $f_0$ is the operating frequency in MHz (note units). This index and derating factor is intended to apply to ultrasound propagating in soft tissue. Note, in this expression, $p_{r}$ is used, as this is the formal definition of MI. Assuming operation in the linear regime, the peak positive and negative pressures are equal, and equal to the pressure amplitude, i.e., $p_{r} = p$, as used in  elsewhere in this guide.

In TUS applications where ultrasound propagates through the skull, a more suitable derating factor may be used to define an alternative index, specific to this application, called the $MI_{tc}$, where the subscript “tc” refers to transcranial applications: 

\begin{equation}
     MI_{tc}=\frac{p_{r, \alpha}}{ \sqrt{f_0} }.
\end{equation}

\noindent In this case, the method used to calculate the \emph{in situ} or derated peak-rarefactional pressure $p_{r, \alpha}$ (for example, a simulation) should be reported. If derating a free field value, $\alpha$, the attenuation coefficient, insertion loss of the skull, and any methods and parameters used for calculation should be reported. If modeling indicates that the target pressure is lower than the peak pressure, then the MI for both the spatial-peak and the target should be given.

\subsection{Thermal Metrics} \label{Sec:thermal}
At least one of the following metrics should be reported.

\begin{enumerate}[(a)]

\item \textbf{Temperature Rise}

In some cases, it may be possible to directly measure the temperature rise in soft tissue due to the applied TUS (for example, using MR thermometry or thermocouples), or to estimate the temperature rise in soft tissue using a thermal simulation. If reporting the temperature rise in soft tissue, the measurement or simulation method should be described as discussed in Sec. \ref{Sec:exposureparams}.

\item \textbf{Thermal Index }
For diagnostic ultrasound imaging, the thermal index ($TI$) provides a standardised indicator related to the potential for thermal bioeffects. $TI$ is a unitless quantity intended to indicate a potential temperature rise in degrees celsius, but doesn’t directly correlate to tissue heating.\footnote{From 60601-2-37 \cite{IEC60601-2-37}: “The TI gives a relative indication of the potential for temperature increase at a specific point along the ultrasound beam. The reason for the term “relative” is that the assumed conditions for heating in tissue are complex such that any single index or model cannot be expected to give the actual increase in temperature for all possible conditions and tissue types. Thus, for a particular beamshape, a TI of 2 represents a higher temperature rise than a TI of 1, but does not necessarily represent a rise of $2^\circ C$.”}
In lieu of a safety standard specifically related to TUS, it is also often reported in TUS studies in relation to limits given in the FDA guidance document \cite{FDAguidance} as evidence of safety. Detailed discussions of these parameters can be found in \cite{IEC62359}. 
Following IEC 62359 \cite{IEC62359}, the relevant thermal index is the bone-at-surface or cranial $TI$ ($TIC$), which is calculated as

\begin{equation}
    TIC = \frac{W}{40 D}.	
\end{equation}

\noindent Here, $W$ is the time-averaged acoustic power of the transducer in free field in mW, and $D$ is the equivalent aperture diameter of the transducer in cm. For transducers in direct contact with the scalp, the equivalent aperture diameter would be taken to be the nominal aperture diameter of the transducer. However, when TUS transducers are applied with some stand off (such as a coupling pad) between the transducer surface and the scalp, the equivalent aperture diameter should be taken as the beam diameter at the scalp. This can be estimated from the distance between the transducer surface and the scalp, and the geometry of the transducer. Therefore, the transducer to scalp distance should be reported.  If reporting TIC, the values for $W$ and $D$ (and the methods used to measure or calculate them) should also be reported.

\item \textbf{Thermal Dose}

The extent of biological changes in tissue resulting from thermal exposure is correlated with the amount of energy absorbed in tissue. However, for thermal energy, it is the temperature to which the tissue is raised, and the duration of the heating that plays the predominant biological role. Sapareto and Dewey \cite{sapareto1984thermal} defined a ‘thermal isoeffective dose’ in terms of cumulative equivalent minutes (CEM) at $43^\circ C$ which allows
conversion of any temperature-time (T-t) combination to the equivalent time for which the reference temperature of $43^\circ C$ must be applied to obtain the same level of thermal damage, given by

\begin{equation}
  CEM = \int_{t=0}^{t=final}{R^{(43-T)}dt},
\end{equation}

\noindent  where $R=0.5$ for $T \ge 43^\circ$ and $R=0.25$ for $T < 43^\circ$. This formula accounts for tissue thermotolerance (e.g. mediated by heat-shock proteins) which occurs during exposure at mild hyperthermic temperatures.

\end{enumerate}

\subsection{Spatial-Peak Pulse-Average Intensity}
\label{Sec:intensity}
Acoustic intensity is a measure of the flow of energy from one point in an acoustic medium to another. The instantaneous acoustic intensity is a vector quantity (it depends on direction) and is given by the product of the acoustic pressure and acoustic particle velocity. For a plane wave, the pressure and particle velocity are related by the characteristic acoustic impedance of the medium (the product of sound speed and density). While plane wave expressions have only limited validity and cannot generally be applied throughout a focused acoustic field \cite{IEC62127}, they are often used to calculate the spatial-peak intensity in a focused field, where the wave is approximately plane. 

Under this assumption, the instantaneous intensity in the direction of the plane wave can be calculated by
\begin{equation}
    I_\mathrm{sp}(t)=\frac{p_\mathrm{sp}(t)^2}{Z}.
\end{equation}

\noindent Here, $p_\mathrm{sp}(t)$ is the time varying acoustic pressure at the location of the spatial-peak, and $Z$ is the characteristic acoustic impedance of the medium, which is approximately $1.5\times 10^6$ Rayls for soft tissue. For a harmonically-varying acoustic wave, assuming that the pulse duration is an integer multiple of the acoustic period (or otherwise long compared to the acoustic period), the intensity time-averaged over the pulse duration (generally called the pulse average intensity) can be calculated using
\begin{equation}
    I_\mathrm{sppa}=\frac{1}{PD~Z} \int_{0}^{PD}{p_\mathrm{sp}(t)^2dt},
\end{equation}

\noindent which, for long pulses without ramping, reduces to
\begin{equation}
    I_\mathrm{sppa}=\frac{p_\mathrm{sp}^2}{2Z},
\end{equation}

\noindent where $p_\mathrm{sp}$ is the spatial-peak pressure amplitude (the amplitude of the sinusoidal pressure signal at the location of the spatial-peak). The $I_\mathrm{sppa}$ should be easily derivable by the reader from the estimated \emph{in situ} pressure and timing, and is therefore an optional reporting parameter. If reporting intensity values, the characteristic acoustic impedance $Z$ used for the calculation must be reported.

\subsection{Spatial-Peak Time-Average Intensities}
\label{Sec:timeav}

The intensity time-averaged over the pulse train or pulse train repetition interval (generally called the time average intensity) can be calculated by 

\begin{equation}
    I_\mathrm{spta}=\frac{1}{T~Z} \int_{0}^{T}{p_\mathrm{sp}(t)^2dt},
\end{equation}

\noindent where $T$ is the time period over which the average is taken. In the case of a pulse train without ramping, this reduces to

\begin{equation}
    I_\mathrm{\text{spta, pulse  train}}= DC_{\text{pulse train}} \; I_\mathrm{sppa}.
\end{equation}

\noindent  Intensity parameters are widely reported in TUS studies. However, as mentioned above, it is important to note that in general, acoustic intensity is a vector quantity. Thus, intensity values calculated in this manner should always be reported alongside (not instead of) acoustic pressure values.

The pulse timing parameters used in TUS frequently include intermittent pulsing, often with different repetition intervals at different pulse levels (see Sec. \ref{Sec:timingparams}). This causes ambiguity in the definition of time-averaged intensity, specifically regarding which time period should be used for averaging. Clearly, if 5 second pulse trains are administered every hour, taking the time-average over the latter period does not give a meaningful quantity. 

Ultimately, the maximum continuous length of time where no TUS is administered included in the $I_\mathrm{spta}$ calculation should be determined based on knowledge of characteristic diffusion times relevant to TUS. However, this is still a topic of active research, thus a recommendation cannot yet be made. For standardised reporting, our current recommendation is therefore to report the time period over which the average is taken. This can be done succinctly using the pulse level nomenclature introduced in Sec. \ref{Sec:timingparams}, e.g. $I_\text{spta, pulse  train}$ and  $I_\text{spta, pulse train repeat}$.

\subsection{Neuromodulation Dose Parameters}

 For TUS, the precise mechanisms through which ultrasound affects brain function are only just beginning to be understood. This means it is not yet possible to define or adopt appropriate dose parameters \cite{shaw2015}.


\section{Summary}

In this paper, we have suggested a minimal set of reporting guidelines for TUS parameters. In the supplementary material, we provide more details on how to obtain free field measurements. We provide some examples of how to estimate \emph{In Situ} Parameters. We briefly list other aspects of the TUS experiment that fall outside of the scope of this paper but are also important in fully describing a study. We also provide a checklist of parameters to be reported. 

\section*{Acknowledgements}

EM was supported by a UKRI Future Leaders Fellowship (MR/T019166/1), by the EIC Pathfinder project CITRUS (Grant Agreement No. 101071008) funded by the EU Horizon Europe research and innovation programme, and by the Wellcome/EPSRC Centre for Interventional and Surgical Sciences (WEISS) (203145Z/16/Z). JFA is  supported by the Focused Ultrasound Foundation Center of Excellence program (Physics for Medicine Paris). L.V. is supported by the Dutch Research Council (NWO) with a VIDI fellowship (18919). BET was supported in part by the Engineering and Physical Sciences Research Council (EPSRC), UK, and the EIC Pathfinder project CITRUS (Grant Agreement No. 101071008) funded by the EU Horizon Europe research and innovation programme. KBP was funded by NIH R01 NS112152, NIH RF1 MH116977, NIH R01 MH131684.

\section*{Declaration of Interests}

EM has acted as a consultant for BrainBox Ltd. JFA holds 5 patents on transcranial ultrasound and had research support from Insightec within the last 3 years. MS is a consultant to Brainsonix Corp., co-inventor on several patents in tFUS,  and is a member of the US Technical Advisory Group to the IEC responsible for diagnostic ultrasound output regulation. BET is a developer of and holds a financial interest in the commercially available k-Plan treatment planning software. LV is a member of the scientific board of the Brainbox Initiative. KBP has acted as a consultant for Attune Neurosciences and has had research support from MR Instruments and General Electric within the last 3 years.

\bibliographystyle{ieeetr}
\bibliography{references}

\appendix
\counterwithin{figure}{section}

\clearpage

\section*{Supplementary Material}

\section{Linearity Assumption}
\label{Sec:App_linearity}
We make the assumption that the ultrasound fields that this guide applies to are linear. Our rationale for assuming so is that the pressure values currently under investigation for the vast majority of transcranial ultrasound stimulation studies are low enough to be in the linear regime. 

We note that linearity means that all or almost all of the energy is at the operating frequency, with minimal energy contained in the harmonics.  Energy transferred to harmonic frequencies during nonlinear propagation of ultrasound results in an asymmetric pressure waveform, where the peak positive pressure becomes greater than the peak negative pressure. 
To verify linearity, the difference between the two pressure values can be examined . This can (and should) be checked by examining the frequency spectrum of the measured pressure waveforms and ensuring that e.g. the amplitude of the second harmonic is $\lesssim$5\% of the amplitude of the fundamental frequency, or the difference between the peak positive and peak negative pressures is not more than $\sim10$\%. 

In our measurements, for a focused ultrasound transducer with a 64 mm radius of curvature and 64 mm aperture diameter operating in water at 500 kHz, the peak positive and negative pressure differ by $>$10\% (indicating that the field is still quasi-linear) only when the pressure amplitude is $>$2 MPa. This corresponds to a Mechanical Index (MI) of $>$2.5. For the same transducer operating at 250 kHz, the peak positive and peak negative pressure differ by $>$10\% only when the pressure amplitude is $>$3 MPa, for which the MI is 6. 

When the field is linear, the size and position of the focal region for a given focal setting can be considered to remain constant between output levels. In addition, this assumption simplifies the calculation of several derived parameters in Section \ref{Sec:derivedparams}. For avoidance of doubt, throughout the guide, the term \emph{pressure amplitude} will be used to describe the amplitude of a sinusoidally varying pressure signal in steady-state under linear conditions, where the pressure amplitude is equal to the peak positive and negative pressures, and half the peak-to-peak pressure (see Fig. \ref{fig:timing}, Fig. \ref{fig:A1} for further details).

\section{Measurement of Free Field Acoustic Pressure Parameters}
\label{AppA}
\subsection{Guidance on Performing Measurements of Acoustic Pressure}
Measurements of acoustic pressure fields are often made using a hydrophone scanned through the field using an automated scanning tank filled with deionised and degassed water. Measurement procedures and the choice of suitable sensors (usually hydrophones) and scanning equipment is specified in detail in international standards for the measurement of medical ultrasonic fields \cite{IEC62127}, with further specific definitions and measurement methods specifically for focusing transducers specified in \cite{IEC61828}. Other sources of guidance and examples of good practice for hydrophone measurements of acoustic pressure fields can be found in the literature \cite{Harris2022, Martin2019, OutputMeas}, and may be more accessible to readers of this document. Important factors to consider in order to reduce uncertainty in measurement include choice of hydrophone, alignment procedures, spatial sampling, and signal acquisition and processing. 

\subsection{Choice of Hydrophone}
Measurements require a suitable sensor which has adequate sensitivity while minimising the effects of spatial averaging, is stable over time, and for the purposes of acquiring absolute pressure measurements, has a calibrated sensitivity (frequency dependent or at least at the frequency of interest). In general, the diameter or width of the hydrophone sensitive element should be less than one quarter of the wavelength of the acoustic field at the frequency of interest. In practice, at the low frequencies ($<$ 1 MHz) often employed in TUS, the effective size of the hydrophone sensitive area can be significantly larger than the nominal size, increasing the effects of spatial averaging.  For example, at 500 kHz, the wavelength of sound in water is $\sim$3 mm, so a hydrophone element diameter of less than 0.75 mm should be used. A hydrophone with a nominal element diameter of 0.5 mm may be suitable, although it should be noted that the effective element size may be significantly larger so it is strongly advised to choose a smaller element size \cite{Martin2019, Wear2022}. If spatial averaging is unavoidable and likely to be significant, corrections may be applied to the pressure measurements to compensate \cite{Wear2022corr}. 

Most commercially available hydrophones can be expected to maintain stable sensitivity provided they are handled carefully and well maintained (i.e. the sensitive element should not be touched, they should not be exposed to pressures above the range of safe operation, they should be rinsed with DI water after use in tap water or other liquids to avoid build up of deposits on the element surface), and in most cases a soaking period of at least an hour should be allowed before measurements are made. Calibrated sensitivity at the frequency of interest traceable to international pressure standards can usually be provided by the hydrophone manufacturer, or obtained directly from a national measurement institute. Calibration at frequencies below 1 MHz may have to be specifically requested.

\subsection{Uncertainty on Hydrophone Measurements}
\label{Sec:UC}

All hydrophone calibrations have an uncertainty which varies with frequency, and calibration method. Hydrophone calibrations obtained from a national measurement institute (NMI) such as the National Physical Laboratory in the UK have a typical uncertainty of 9\% in the range 250 kHz to 1 MHz. The uncertainty on hydrophone calibrations obtained from hydrophone manufacturers is typically approximately double that associated with NMI calibrations, up to 20\% or more.

Other sources of systematic uncertainty include spatial averaging, which can become significant for large hydrophones in focused fields. Random uncertainty also arises from various sources including positioning, alignment and electrical fluctuations in amplifiers and source impedance. Despite the relatively high uncertainties, it has been demonstrated that reproducibility and repeatability of pressure measurements can be within 10\% \cite{Martin2019}. 
It should also be noted that the uncertainty on pressure measurements will propagate to derived quantities. As intensity is proportional to the square of pressure, this results in a doubling of the uncertainty from pressure to intensity.   

When comparing multiple pressure measurements of the same field, for example where sequential measurements are performed at intervals, or measurements are obtained from different sources, agreement is demonstrated if two measurements differ by less than the uncertainty on the measurements. Where multiple measurements agree to within the uncertainty, then an average value can be reported. If the difference is larger, it may indicate some change in performance of the device or error in the measurement process. 

\subsection{Signal Parameters and Acquisition}
To achieve free field conditions under which the effect of reflections and interference are suppressed during measurements, the pulse parameters used for measurement of free field pressure parameters are usually different from those used during the study itself. The pulse length and timing of the signal waveform acquisition is chosen so that the measured field represents the field that would exist if the transducer was radiating into water continuously or with the comparatively long pulses used in TUS studies \cite{international1iec} . These timing conditions together with a water bath of sufficient volume or with a suitable absorbing lining ensures that interference from acoustic reflections is reduced or eliminated.

\begin{figure}
    \centering
    \includegraphics[width=\textwidth]{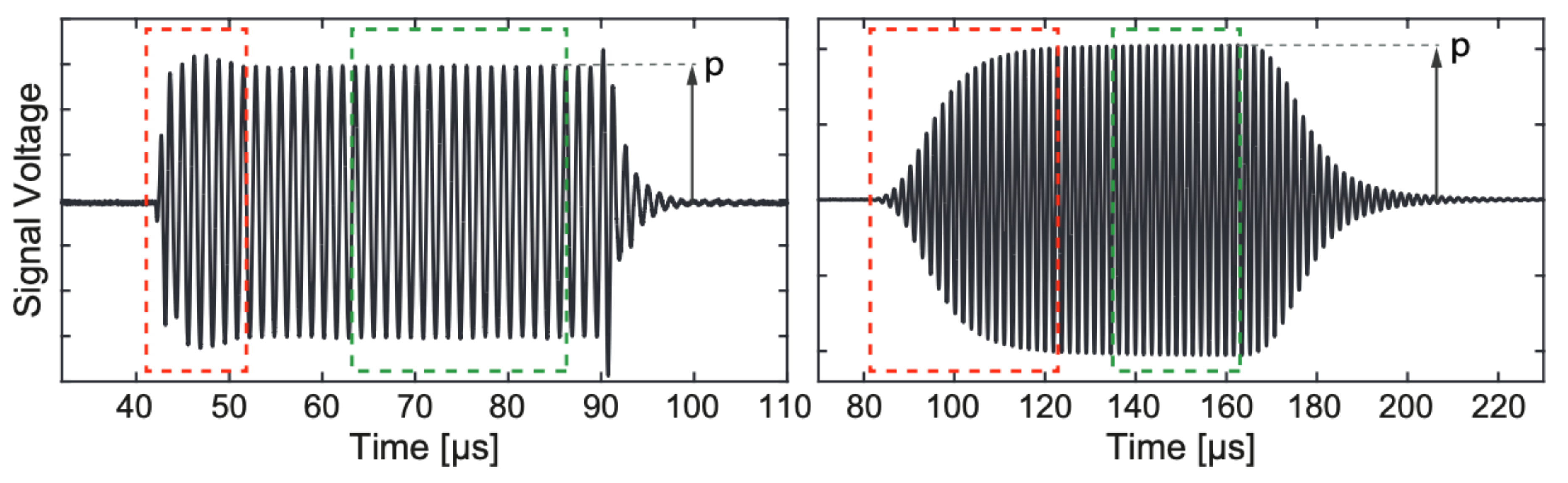}
    \caption{Example focal waveforms measured at the position of spatial-peak pressure under free field conditions. The region within the red dashed box is the transducer ring up period before the waveform reaches a steady amplitude. The region within the green dotted box shows the steady part of the waveform which should be used to calculate the pressure amplitude, denoted as $p$. The figure on the left shows a measurement from a transducer with a broad bandwidth (short ring-up time), and on the right a transducer with a narrow bandwidth (long ring-up time).    }
    \label{fig:A1}
\end{figure}

These quasi steady state conditions are achieved in a time window which starts after sound emitted from all parts of the transducer has reached all measurement locations, with additional time to allow for the ring-up time of the transducer, and ends before reflections from the measurement equipment reach the hydrophone (e.g. sound which is reflected from the hydrophone back to the source and then back to the hydrophone). The pulse duration should be chosen to be long enough to include some or all of this period and to ensure that there is a period of the waveform which has a consistent amplitude. The pulse duration should be short enough to avoid overlap of electrical pick up from transducer drive voltage with the acquisition window, and reflections from other parts of the measurement set up. Most transducers have a ring up time of at least a few cycles at the beginning of the pulse. This is the length of time it takes for the pulse amplitude to reach a steady value. Depending on the bandwidth of the transducer, this varies from a few cycles to 10s of cycles. Measurements of pressure amplitude should be made using part of the waveform which occurs after this period (see Fig. \ref{fig:A1}).

Voltage signals from the hydrophone are usually acquired using an oscilloscope. The temporal sample rate should be sufficient to accurately represent the waveforms, with a recommended sample rate of at least 20 times the ultrasound frequency of interest \cite{IEC62127-2}. The oscilloscope should be triggered using the trigger output from the drive system. This should provide a stable acquisition and provide the delay time between emission of the pulse from the transducer, and arrival of the acoustic signal at the hydrophone. This can be used to calculate the axial position of the hydrophone relative to the transducer, given knowledge of the sound speed of water at the measurement temperature. It is recommended that this axial position calculation is performed at the focus of the field where the arrival time of the pulse can be most easily determined. 

\subsection{Alignment and Spatial Sampling}
It is recommended that hydrophone measurements of the spatial distribution of acoustic pressure are performed in deionised and degassed water of known temperature, using an automated scanning tank with which the hydrophone can be positioned reproducibly to within 0.05 mm along three orthogonal directions. 
Following a rough alignment by eye, the hydrophone should be aligned to the beam axis by alternately scanning along the two lateral directions through the centre of the -6 dB focal region until the position of the centre is consistent to within approximately 100 $\mu$m. Finding the midpoint using the centre of mass of the -6 dB region rather than the peak value is recommended, since it is less susceptible to noise and fluctuations. 

The transducer may be mounted in a fixed geometry using a mount that ensures that the beam axis is aligned with the parallel tank scanning axis, this should be verified. It is recommended that before acquiring pressure profile measurements, the hydrophone should be aligned to the spatial peak pressure, to ensure that all line measurements pass through that through the focus. Where the transducer mount geometry is not fixed and adjustment of the angular position is possible, the transducer beam axis should be aligned with the parallel scanning axis. This can be achieved by aligning the hydrophone with the beam axis at two axial distances to determine any angular offset, and correcting the angular tilt and rotation of the transducer accordingly. Alternatively, where scans can be performed along arbitrary directions, the scan coordinates should follow the beam axis.

Measurements of the field should be made with adequate spatial sampling, of less than half a wavelength or finer in order to accurately capture spatial features and peak pressure values. Rather than acquiring point measurements, it is recommended that orthogonal line scans are acquired along lines which pass through the location of the spatial-peak pressure. Once the hydrophone is aligned with the transducer beam axis, an axial line scan should be performed to identify the axial position of the spatial-peak pressure. 

The axial position relative to the transducer can be estimated using the time of flight of the acoustic signal. However, there may be some error in this distance where there are e.g. time delays between the trigger signal and generation of the acoustic output due to the system electronics.  Additionally, where the transducer construction includes lenses or coupling media, the measurement coordinates are often defined relative to a reference plane which is not the origin of the acoustic signal. In this case, a physical measurement of the reference plane to hydrophone distance should be made. This can be done using e.g. a force sensor, dummy hydrophone, or offset block.   

\subsection{Signal Processing}
Hydrophones output a voltage signal which must be converted to pressure using the known hydrophone sensitivity. Some signal processing steps should be taken to ensure that errors are not introduced during this conversion. For linear fields, the peak negative pressure, peak positive pressure and the pressure amplitude can be expected to be equivalent. Peak negative pressure is usually given as a quantity linked to mechanical effects. The values can be obtained by several methods, but it is recommended that a method is chosen which reduces the impact of noise and fluctuation of the signal amplitude throughout the acquisition window.

For quasi steady state signals, the pressure amplitude can be acquired by performing a fast Fourier transform (FFT) on a measured waveform which has been trimmed to include a whole number of cycles from a steady part of the waveform, then extracting the magnitude of the FFT at the frequency of interest. The length (number of samples) in the voltage signal should be large enough to provide adequate sampling in frequency space so that the amplitude at the frequency of interest can be obtained. 
Alternatively, the minimum value of each cycle from a steady part of the acquired waveform can be obtained and averaged. 
The peak negative voltage or voltage amplitude can then be converted to pressure by dividing by the hydrophone sensitivity at the frequency of interest given in units of V/Pa, or by deconvolution of the hydrophone response from the voltage waveform.

\subsection{Scaling Spatial-Peak Pressure Measurements by Drive Level}
The free field spatial-peak pressure amplitude at each of the study output level settings should be reported. The pressure can be obtained either directly from a measurement with the transducer operated at the study output level and focal setting, but in practice, it may be simpler and more practical to establish a scaling from the measurement output level, or the output level given in the manufacturer’s test report, to the study output level(s) for a given focal setting.

While the acoustic pressure field is linear, the output pressure of most systems is expected to scale linearly with the voltage applied to the transducer. So for example, focal pressure amplitude should scale linearly with drive voltage amplitude. Correspondingly, if the output setting used is power or intensity, since this is proportional to pressure squared, the spatial-peak pressure should scale linearly with the square root of the power or intensity setting. It is assumed that at the range of output levels used for TUS, the location and volume of the focal region will remain constant provided no changes are made to the focal settings.

The relationship between output level setting and spatial-peak pressure should be verified if possible by making several measurements of the spatial-peak pressure at different output level settings for a given focal setting. A line fitted to the measurement points can then be used to obtain the pressure for any output level setting later used in studies. Note that if the focal position setting is changed, a different scaling is needed. 

\subsection{Decibels and Focal Size Quantities} \label{dB}
\begin{figure}
    \centering
    \includegraphics[width=\textwidth]{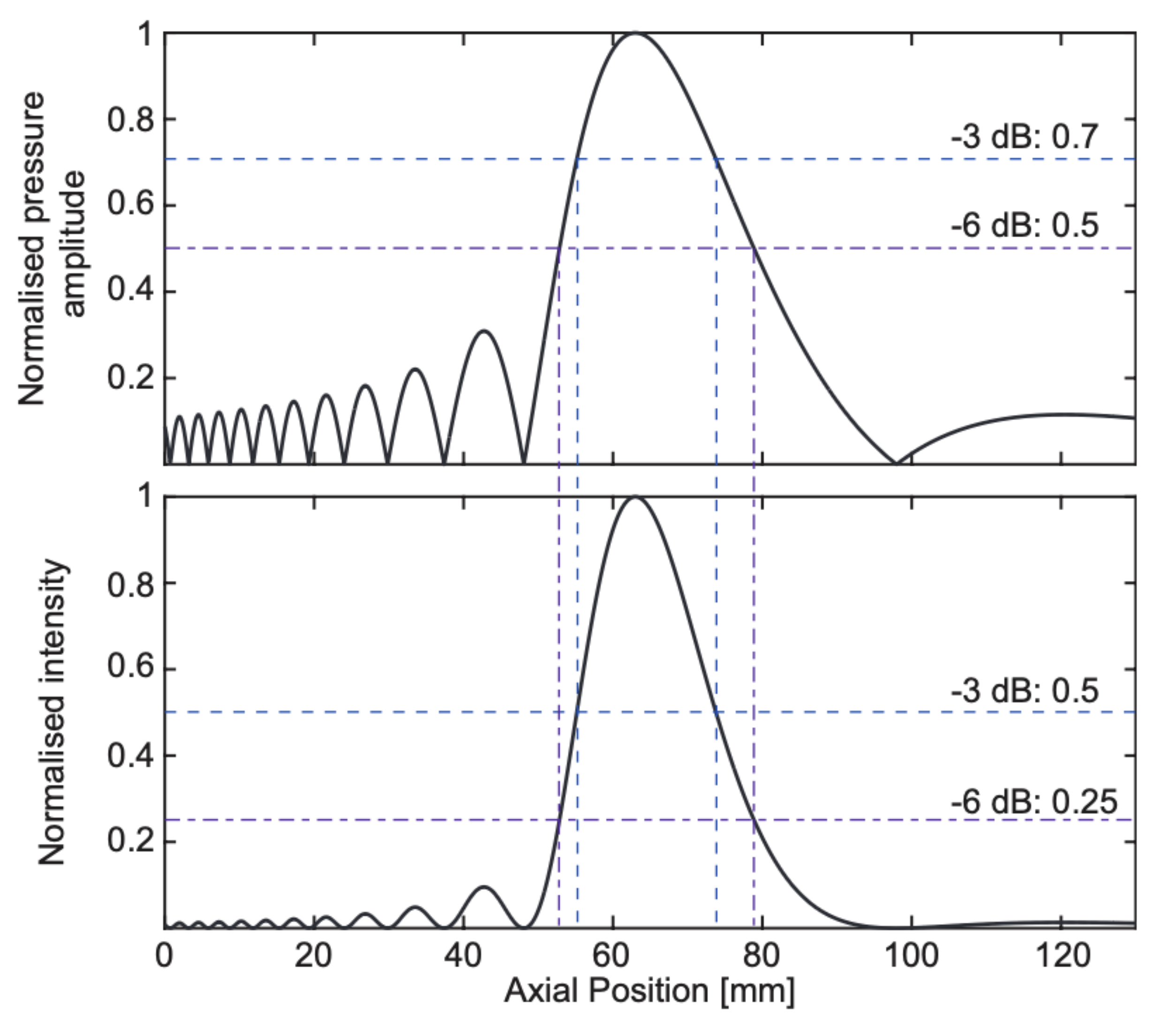}
    \caption{Axial profiles of normalised pressure amplitude (top) and corresponding normalised intensity (bottom), showing the -3 dB and -6 dB axial focal lengths which correspond to different fractional thresholds in pressure and intensity.}
    \label{fig:A2}
\end{figure}

In acoustics, decibels are used to describe the loudness of sound on a logarithmic scale, as the range of pressure amplitudes in the human audible range is large, similarly for the range of pressure amplitudes in the transmitted and received pulses in ultrasound imaging. When used in ultrasound field characterisation, they are often used to describe a change or difference in pressure or intensity. These differences in units of decibels [dB] are calculated from the log to the base 10 of the ratio of the pressure or intensity at the point of interest, $p$, to the reference pressure or intensity, $p_{ref}$: 

\begin{equation}
     10\ \log_{10}\left(\frac{I}{I_{ref}}\right) = 10\ \log_{10}\left(\frac{p^2}{p^2_{ref}}\right) = 20\ \log_{10}\left(\frac{p}{p_{ref}}\right) \ [dB]
\end{equation}

In the context of determining focal length and width values, the reference pressure or intensity is the spatial-peak value.
The same focal length value can be calculated from either the pressure or intensity profiles in terms of dB, with the appropriate fractional threshold (see Fig. \ref{fig:A2}).  The full width half maximum of the intensity profile is the -3 dB region, but in terms of pressure, this -3 dB region corresponds to the region over which the pressure exceeds 70\% of the maximum value. 
Note that these quantities are useful for comparison of fields in water or for looking at focal distortion in situ, for example, but since they are relative quantities, they do not indicate the region of potential effect, since this is likely to correspond to an absolute pressure threshold.

\subsection{Obtaining Pressure Quantities from Manufacturer's Report}
It is expected that the manufacturer should provide some conversion between the setting used to control the output level and the spatial-peak pressure for at least a subset of focal settings. It may not be possible to obtain all required parameters for all possible focal settings from the manufacturers report, but all parameters that can be obtained should be reported. For example, if the position setting refers to the centre of the focal region rather than the location of the spatial-peak pressure, then it may not be possible to report both for every focal setting used. The difference between these positions will vary depending on the transducer parameters and focal settings, it may be as much as 10 or 20 mm.

\section{Reporting Average Parameters}
\label{Sec:av_params}
In some cases where a study includes a large number of participants, each with personalised ultrasound parameters, it could be impractical to report the free field parameters, drive system settings, and estimated \emph{in situ} exposure parameters under each condition used. In this situation, it is appropriate to report the mean, minimum and maximum values. 

In the simplest case, the same drive system settings and free field pressure parameters may be applied for all participants, but individualised \emph{in situ} exposure parameters obtained from simulations using the participants anatomy. In this case, it is reasonable to report the mean, minimum and maximum of the estimated \emph{in situ} exposure parameters across all participants.

In another the case, the drive system settings may be adjusted with aim of delivering the same ultrasound pressure to the same brain region across subjects. In this case, the range of drive system settings used should be reported. The mean, minimum and maximum of the free field spatial-peak pressure amplitude and size of the focal volume should be reported, and the range of the positions of the free field spatial-peak pressure amplitude. 

Where an array transducer is used with steering and aberration correction to obtain the same \emph{in situ} spatial-peak pressure amplitude across subjects, free field measurements made under the exact conditions are not relevant and would be impractical to perform for every condition. In this case, measurements of the free field pressure parameters should be made at the geometric focus or central position, and over the steering range employed during the study. Free field spatial-peak pressure amplitudes and size of the focal volume measured across the steering range relative to the geometric focus or central position will illustrate the steering and focusing performance of the array. The range of drive system settings used should also be reported. The range of \emph{in situ} size of the focal volume obtained from simulations used to perform the aberration correction should be reported, since these may vary depending on position and skull anatomy.  

\section{Example Estimates of \emph{In Situ} Parameters}
\label{Sec:ex_insitu}

\subsection{Estimated \emph{In Situ} Pressure Amplitude}
This example adapted from Deffieux 2013 \cite{deffieux2013low} reports attenuation from only skull in the following way: ``\emph{The pressure amplitude at focus was set to 0.6 MPa, as measured in free water with a heterodyne interferometer. Skull transmission was estimated on a clean and degassed primate skull specimen (Macaca mulatta skull) at seven different locations and was found to be 58\% ± 8\% (derating factor of -4.7 dB). This allowed us to estimate the derated spatial-peak pressure at 0.35 MPa in the brain of the monkeys.}"

\subsection{Estimate of \emph{In Situ} Mechanical Index}
The following example calculation uses values taken from Johnstone \textit{et al.} \cite{johnstone}: ``\emph{The $MI$ was 0.44, calculated as follows. The spatial-peak pressure amplitude measured in water was 700 kPa. Assuming a constant derating factor of -9.8 dB (the average insertion loss of the skull at 270 kHz measured in \cite{albelda2019experimental}), the derated spatial-peak pressure amplitude was 230 kPa. The operating frequency was 270 kHz. The $MI_\text{tc}$ is then 0.23 / sqrt(0.27) = 0.44.}

\subsection{Thermal Metrics}
The following example calculation uses values taken from Johnstone \textit{et al.} \cite{johnstone}: ``\emph{The TIC was 0.48, calculated as follows. The electrical power was 4.8 W. Assuming a nominal electrical efficiency of 85\%, this gives an acoustic power of 4.1 W. The minimum pulse train repetition interval was 5 seconds, giving an overall duty cycle of 3\% (150 ms on every 5 s). This gives a time-averaged power of 0.12 W. The nominal aperture diameter was 64 mm. The TIC is then (1000 * 4.8 * 0.85 * 0.03) / (40 * 6.4) = 0.48 .}"

\label{AppB}
\section{Additional Experimental Reporting}
\label{Sec:additionalparams}
There are many other aspects to the TUS experiment that critically impact the experiment and should be reported. Due to the large number of different measurement conditions, equipment, and experimental setups, it is not possible to be prescriptive about how these should be reported. Here, we briefly list other aspects of the TUS experiment that fall outside of the scope of this paper but are also important in fully describing a study.

\begin{itemize}[itemsep=-0.1em]
    \item Hair preparation
    \item Subject positioning (e.g., sitting or lying) and fixation (e.g., chin rest)
    \item Transducer alignment and fixation
    \item Neuronavigation equipment and procedure
    \item Auditory perception
    \item Masking and earplugs
    \item Sham conditions and blinding
    \item Environmental conditions (e.g., background noise and light levels)
    \item Additional measurement equipment
    \item Subject experience, side effects, and adverse events
    \item Any safety and electromagnetic compatibility tests performed 
\end{itemize}

\clearpage

\section{Checklist} \label{Sec:checklist}
\newcommand{\checkbox}{\item[$\square$]}

Report these parameters for all devices and settings used. Further details for each parameter are given in the referenced section.

\subsection*{Transducer and Drive System Description}
\begin{itemize}[label={}, itemsep=-0.1em]
  \checkbox Transducer manufacturer and model number (Sec. \ref{Sec:tx_desc}) 
  \checkbox Transducer centre frequency (Sec. \ref{Sec:tx_desc})
  \checkbox Transducer geometry (e.g., radius of curvature and aperture diameter) (Sec. \ref{Sec:tx_desc})
  \checkbox Drive system components, including manufacturer and model number (e.g., signal generator and amplifier or integrated driving system) (Sec. \ref{Sec:drive_system_desc})
   
\end{itemize}

\subsection*{Drive System Settings (Sec. \ref{Sec:drive_system_settings}) }
\begin{itemize}[label={}, itemsep=-0.1em]
  \checkbox Operating frequency 
  \checkbox Output level settings 
  \checkbox Focal position settings
  \checkbox Description of transducer coupling method

\end{itemize}

\subsection*{Free Field Acoustic Parameters (Sec. \ref{Sec:ff_param})}
\begin{itemize}[label={}, itemsep=-0.1em]
  \checkbox Reference position for measurements
  \checkbox Spatial-peak pressure amplitude
  \checkbox Position of spatial-peak pressure amplitude (relative to reference position)
  \checkbox Size of focal volume (-3 dB and -6 dB axial lengths and lateral widths)
  \checkbox Position of centre of focal volume (centre of -3 dB relative to reference position)
  \checkbox Description of how free field parameters were obtained
\end{itemize}

\subsection*{Pulse Timing Parameters (Sec. \ref{Sec:timingparams})}
\begin{itemize}[label={}, itemsep=-0.1em]
  \checkbox Pulse timing table
\end{itemize}

\subsection*{\emph{In Situ} Estimates of Exposure Parameters}
\begin{itemize}[label={}, itemsep=-0.1em]
  \checkbox Estimated \emph{in situ} spatial-peak pressure amplitude (Sec. \ref{Sec:exposureparams})
    \checkbox Estimated \emph{in situ} pressure amplitude at the target (Sec. \ref{Sec:exposureparams})
  \checkbox Estimated \emph{in situ} mechanical index (Sec. \ref{Sec:MI})
  \checkbox One of the following thermal metrics: temperature rise, thermal index, or thermal dose (Sec. \ref{Sec:thermal})
  \checkbox Description of how \emph{in situ} estimates were obtained  (Sec. \ref{Sec:exposureparams})
\end{itemize}

\subsection*{Intensity Parameters (Optional)}
\begin{itemize}[label={}, itemsep=-0.1em]
  \checkbox Spatial-peak pulse-average intensity (Sec. \ref{Sec:intensity})
  \checkbox Spatial-peak time-average intensities (Sec. \ref{Sec:timeav})
  \checkbox The acoustic impedance used for the conversion (Sec. \ref{Sec:intensity})
\end{itemize}

\label{Sec:App_checklist}


\end{document}